# Lattice Vibrational Modes and their Frequency Shifts in Semiconductor Nanowires

*Li Yang [†§] and M. Y. Chou [†‡\*]*

[†]School of Physics, Georgia Institute of Technology, Atlanta, Georgia 30332-0430, USA

[§]Department of Physics, Washington University, St. Louis, Missouri, 63130, USA

[‡]Institute of Atomic and Molecular Sciences, Academia Sinica, Taipei 10617, Taiwan

Abstract: We have performed first-principles calculations to study the lattice vibrational modes and their Raman activities in silicon nanowires (SiNWs). Two types of characteristic vibrational modes are examined: high-frequency optical modes and low-frequency confined modes. Their frequencies have opposite size dependence with a red shift for the optical modes and a blue shift for the confined modes as the diameter of SiNWs decreases. In addition, our calculations show that these vibrational modes can be detected by Raman scattering measurements, providing an efficient way to estimate the size of SiNWs.

The lattice vibrations in nanostructures are modified by both the surface and quantum confinement effects compared with their bulk counterpart. Among various nanostructures, silicon nanowires (SiNWs) are of particular interest because of the fundamental role of silicon in semiconductor physics and broad device applications.[1–4] Recent experiments have discovered a number of specific features in SiNWs related to phonons, such as a low thermal conductivity,[5] unusual electron-phonon interactions,[6] and asymmetry of the Raman spectra of optical modes.[7] In particular, SiNWs have promising thermoelectric properties with important potential for energy applications.[8–10]

Many of previous theoretical investigations in this area employed empirical models with parameters obtained from the bulk. Several unique phonon properties of nanostructures have been identified;[11–14] however, it is difficult to assess the accuracy of these model calculations. More recent first-principles calculations revealed interesting properties of the phonon modes of SiNWs in the low-frequency regime, such as the radial breathing mode (RBM)[15] and a significant reduction of group velocity.[16] Those confined modes are expected to provide essential structural information, but the RBM and its Raman activity will inevitably decay for thicker NWs because of weaker quantum confinement there. Therefore, in this study we focus on first-principles calculations of the vibrational modes for the whole frequency range. It turns out that the optical modes may provide a better way to estimate the size of NWs.

In this paper, we report first-principles calculations of the lattice vibrational modes of SiNWs along the [110] direction. We focus on the modes at the $\Gamma$ point because they are accessible by Raman scattering measurements. In addition to showing the low-frequency confined modes, e.g., the RBM, and their physical origins, our calculation reveals a novel splitting and red shift of high-frequency optical modes resulted from the surface effect. More importantly, the corresponding Raman activity of these high-frequency optical modes will not decay as the diameter of SiNWs increases, making them a convenient method to detect the size of various semiconducting NWs.

We perform the calculations for the SiNWs using density functional theory (DFT) within the local density approximation (LDA) as implemented in the Quantum ESPRESSO package.[17] The surface dangling bonds are passivated by hydrogen atoms, and all calculations are done in a supercell arrangement[18] using a plane-wave basis and norm-conserving pseudopotentials.[19] The vacuum region between neighboring NWs is chosen to be about ~ 1 nm in order to avoid spurious interactions, and the plane-wave energy cutoff is 18 Ry. For the Brillouin zone (BZ) integration we use a 1x1x8 sampling grid. For phonon calculations, we use the linear response approach to obtain the lattice vibrational modes and their frequencies.[20–23] We have investigated three SiNWs with the diameter ranging from 1.2 to 2.2 nm. The force and stress are fully relaxed.

Figure 1 shows the relaxed structure of a SiNW along the [110] direction with a diameter of 1.2 nm. For this particular orientation, the dangling bonds on the [1$\bar{1}$0] surface have to be manually titled in order to achieve a low-energy configuration, which is necessary to avoid unstable modes as pointed out in a previous study of the silicon surface.[24]

We have first calculated the lattice vibrational modes and their frequencies. In Figure 2, we present the calculated density of modes (DOM) at the $\Gamma$ point of the SiNWs. For comparison, the corresponding DOM of bulk silicon is shown in Figure 2(a), whose two main features are found at 4 THz and around 15 THz, respectively. The results for SiNWs of three different diameters are presented in Figure 2(b)-(d). The vibrational modes associated with Si-H bonds are not shown because they are not of interest in the present study and their frequencies are much higher (> 18 THz). As shown in Figure 1, SiNWs preserve the bulk tetrahedral structure very well. Therefore, their DOMs are overall quite similar to each other and approach that of bulk silicon as diameter increases, as shown in Figure 2. On the other hand, since the structure of SiNWs has a lower symmetry than that of bulk silicon, the degeneracy of many phonon modes is lifted, resulting in a wider distribution in the DOM of SiNWs. In addition, quantum confinement can introduce modes that do not exist in the bulk, further broadening the DOM features in SiNWs. The broadening of DOMs has been observed in various nanostructures in recent experiments.[25–27]

Since quantum confinement plays an extremely important role to decide the electrical and thermal properties of semiconducting nanostructures, we first focus on those vibrational modes resulting from confinement.[28] According to our calculations, most of such confined modes are located within the low-frequency regime (roughly below 8 THz in Figure 2). In Figure 3, we present representative confined modes that have special symmetries and are different from the lattice vibrational modes in bulk silicon. One type of them is the RBM shown in Figure 3 (a) and (b), with atomic displacements along the radial direction only. Because of the boundary condition, there exists an amplitude modulation along the radial direction, generating a nodal structure as shown in Figure 3 (b). Other types of confined modes are shown in Figure 3 (c) and (d). Figure 3 (c) illustrates a rotational mode whose atomic displacements have mainly the angular component. The vibrational mode in Figure 3 (d) has both the radial and angular components, and each of these two components has its own spatial modulation.

Because the geometry of SiNWs is close to that of a cylinder, an elastic media model with a cylindrical boundary condition is appropriate to describe the features of the confined modes. The simplest form to the elastic wave function is a standard first-order Bessel equation and its solution is the first-order Bessel function $J_1$. For the free boundary condition that requires the curvature of $J_1$ to be zero, it is easy to find that the wave number $k$ has to be proportional to $1/d$ ($d$ is the diameter of SiNWs). This means that the frequency ($\omega$) of the RBM is proportional to the inverse of the diameter of SiNWs since $k = \omega/c$ ($c$ is the sound velocity). In Figure 4 (a), we have plotted the calculated frequency of the RBM of SiNWs, and fit the first-principles data using a $(1/d)^\alpha$ function. The fitted value of $\alpha$ is around 1.2. Other types of collective modes, e.g., those in Figure 3 (c) and (d), can also be qualitatively described by the elastic wave model with both the radial and angular components.

Because of the size dependency of these confined modes, they are usually regarded to be of practical importance to estimate the size of NWs through Raman scattering measurements.[29–31] Following the approach outlined in Ref.,[23] we have calculated the Raman spectrum of the narrowest SiNW ($d$ ~ 1.2 nm) and the result is shown in Figure 2 (e). Because of the highly anisotropic nature of SiNWs, we have to perform the space average and include the depolarization effect into the Raman spectrum.[32] It is clear from Figure 2 (e) that the RBM and many other confined modes shown in Figure 3 are Raman-active. In particular, the RBM is strongly active in this narrow SiNW ($d \sim 1.2$ nm), so it can be used as an efficient quantity to estimate the size of narrow NWs.[33]

However, it is expected that the Raman activity of the RBM and other confined modes will decay and gradually disappear when the size of SiNWs increases, since only the prominent peak associated with the high-frequency optical modes appears in the first-order Raman spectrum of bulk silicon. Therefore, the high-frequency optical modes would be a better candidate to provide a consistent measure of the size of the SiNWs over a large range.

In the following we focus on the three high-frequency optical modes, which are degenerate in the perfect bulk silicon crystal. The reduced symmetry in SiNWs split them into two groups: two transverse optical modes (TO1 and TO2) with their atomic displacements perpendicular to the axis, and one longitudinal optical mode (LO) with its atomic displacements along the axis. The details of the displacements of these optical modes are plotted in Figure 5. The frequencies of these optical modes are marked in Figure 2; the two TO modes are nearly degenerate and substantially separated from the LO mode. The separation increases as the diameter of SiNWs decreases. At the same time, the frequencies of all these three optical modes decrease with decreasing diameter of SiNWs, as shown in Figure 2. This red shift of optical modes can be attributed to the loss of Si-Si bonds on the surface of SiNWs, the number of which scales as the length of the circumference. The fraction of atoms at the surface then scales as $d/d^2 \approx 1/d$. Therefore, we expect the frequency red shift to scale approximately as $1/d$. In Figure 4 (b), we have plotted the frequency shifts of these optical modes and find that a $(1/d)^\alpha$ function fits these shifts reasonably well with the fitted $\alpha$ being around 1.2-1.5.

It is interesting that the LO mode shows a larger red shift than the other two TO modes as shown in Figure 2 and Figure 4 (b). This can be understood by the following: the surface of SiNWs terminates a significant number of surface Si-Si bonds that mainly affect the bond-stretch motion for the LO mode and the bond-bending motion for the TO modes, respectively, as illustrated by atomic displacements in Figure 5. Since the stretching stiffness is bigger than the bending stiffness, the influence on the LO mode by this surface termination is more significant than on the TO modes, resulting in a larger red shift for the LO mode.

The red shift of the optical phonon modes in SiNWs is consistent with the behavior found in Si nanocrystals. Khoo et al. [34] have performed first-principles density functional calculations to investigate the size dependence of the Raman spectra of Si nanocrystals. Although the surface Si-Si bond length is reduced, they found that the Raman red shift of the optical phonon mode is mainly due to an increased proportion of under-coordinated atoms present at the nanocrystal surface, leading to a softening of mode frequencies in smaller nanocrystals. In our calculated SiNWs, we have observed a similar change of Si-Si bond length (around 0.5%) when approaching the surface. This supports that the under-coordinated atoms may be a more important factor to decide the red shift of optical modes. These results also indicate that the red shift is much less dependent on the relaxation of the wave-vector selection rule in a confined geometry as proposed previously [35].

The Raman activities of these high-frequency optical modes are presented in Figure 2 (e). Due to the depolarization effect, only transverse modes ("T" modes) are highly active. Even though the red shift and splitting (~ tenths of one THz) are much smaller than those of the confined modes, they are within the resolution of current Raman scattering measurements. Therefore, compared with the RBM, the splitting and red shifts of these high-frequency optical modes provide a better approach to estimate the size of bigger NWs.

We have discussed above two types of vibrational modes: one is high-frequency optical modes and the other is confined modes. Interestingly, although the frequencies of both types of vibrational modes display a size dependence, their variation with respect to the nanowire diameter is opposite to each other. As the diameter of SiNWs decreases, the optical modes show a red shift, while the RBMs show a blue shift. This is due to the different physical origin of these shifts. Since the frequency shift of optical modes arises from the loss of Si-Si bonds on the surface, narrower SiNWs result in "softer" or red-shifted modes. On the other hand, confined modes such as RBMs are the result of quantum confinement, so narrower SiNWs exhibit a larger confinement effect with a larger energy increase.

In conclusion, we have performed first-principles calculations to study lattice vibrational modes and their Raman activities of SiNWs. Two important types of vibrational modes are studied: the high-frequency optical modes and the low-frequency confined modes. The size dependence of their frequencies is also discussed. As the diameter of SiNWs decreases, the frequency of optical modes shows a red shift while that of the RBM shows a blue shift. An elastic wave model for a cylinder is applied to explain the frequency shift of the confined modes. At the same time, the Raman spectrum of SiNWs is calculated. Our study predicts that the unique splitting and red shift of the high-frequency optical modes can be employed to estimate the size of SiNWs.

This work is supported by the US Department of Energy, Office of Basic Energy Sciences, Division of Materials Sciences and Engineering under Award No. DEFG02-97ER45632. Computational resources are provided by the National Energy Research Scientific Computing Center (NERSC).

Figures and captions:

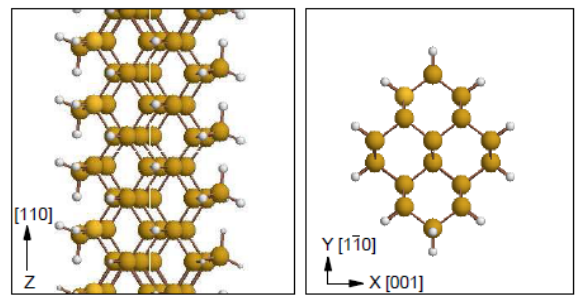

Figure 1: Structure of the smallest SiNW along the [110] direction. The left is the side view and the right is the top view. The atoms on the $[1\bar{1}0]$ surface are titled to stabilize the structure.

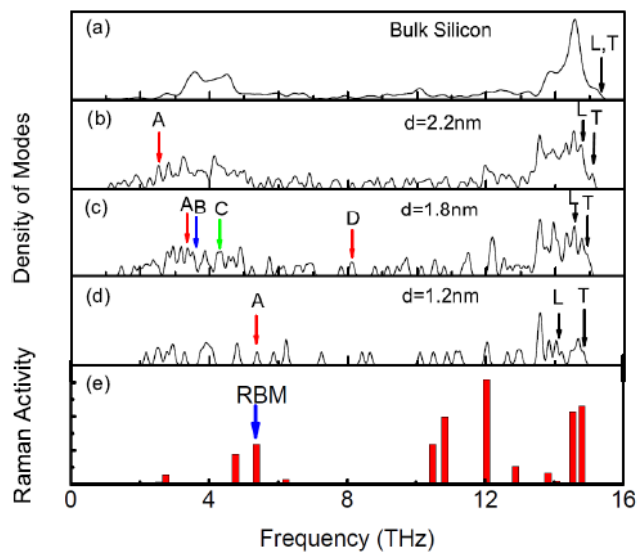

Figure 2: (a)-(d) Density of vibrational modes (DOM) for bulk silicon and three SiNWs with different diameter sizes. L denotes the LO mode and T denotes the TO modes, respectively. A, B, C, and D are the collective (confined) modes shown in Figure 3. A 0.05 THz Gaussian broadening is applied. (e) The Raman activity of the 1.2 nm SiNW with an arbitrary unit and the radial breathing mode (RBM) is marked.

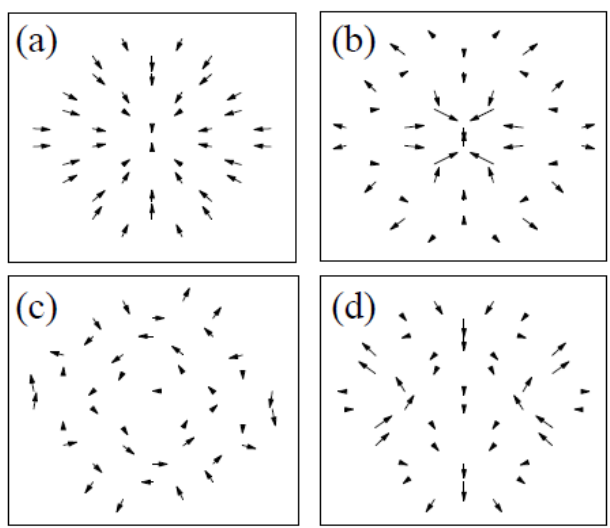

Figure 3: Typical confined modes of a SiNW ($d$ =1.8 nm). They are also marked as A-D in Figure 2 (c).

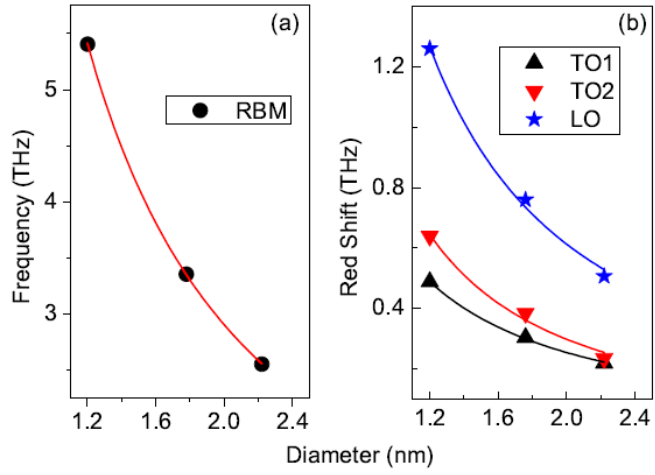

Figure 4: (a) The blue shift of the radial breathing mode (RBM) and (b) the red shift of three high-frequency optical modes. Solid curves ($1/d^\alpha$) are used to fit corresponding data points. $\alpha$ is about 1.2 in (a) and 1.2~1.5 in (b).

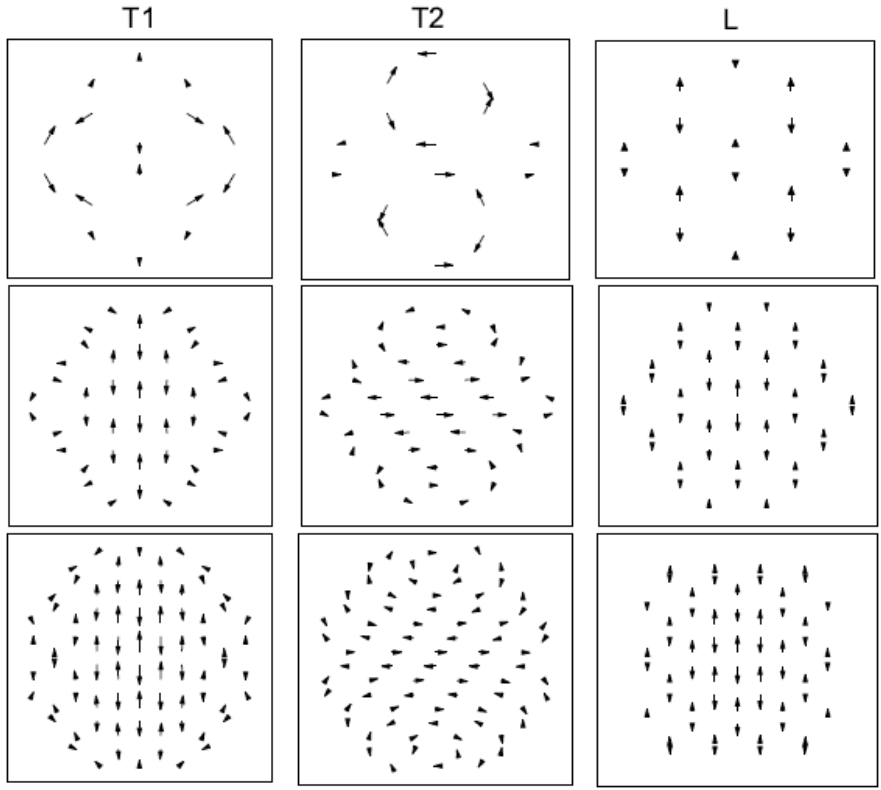

Figure 5: Three high-frequency optical modes in SiNWs. Each column represents a specific type of optical modes but for different SiNWs; L denotes the LO mode and T denotes the TO modes, respectively. For LO modes, atomic vibrations are projected along the y direction. The displacements

are shown for three SiNWs with different sizes. The atomic positions are rescaled in order to compare their displacements.